\documentclass[aps,pra,reprint]{revtex4-2}
\usepackage{graphicx} 
\usepackage{float}
\usepackage{amsmath}
\usepackage[normalem]{ulem}
\usepackage{comment}
\usepackage{csquotes}
\usepackage{multirow}
\usepackage{xcolor}

\usepackage{setspace}
\usepackage[margin=1in]{geometry}

\usepackage[normalem]{ulem}
\newcommand{\stkout}[1]{\ifmmode\text{\sout{\ensuremath{#1}}}\else\sout{#1}\fi}

\usepackage{xcolor}

\begin{document}

\title{Probing microcavity resonance spectra with intracavity emitters
}

\author{C. Koks}
\author{M. P. van Exter}
\affiliation{Huygens-Kamerlingh Onnes Laboratory, Leiden University,
P.O. Box 9504, 2300 RA Leiden, The Netherlands}

\date{\today}

\begin{abstract}
We measure the fluorescence spectrum of broadband emitters in an open optical microcavity with radius of curvature $R=17.7(3)~\mu$m and finesse $F\approx1000$.
This geometry enables a combined measurement of emission spectra versus cavity length, which has several benefits over measurements 
at fixed wavelength or fixed cavity length alone.
We demonstrate the role of the optical penetration depths on the cavity modes and provide practical working equations for its analysis.
Furthermore, we show the ability to measure the coupling of cavity modes within a small scan range of the cavity length.
By measuring these cavity emission spectra as a function of cavity length, we thus obtain a rich and complete picture of the optical microcavity.
\end{abstract}

\maketitle

\section{Introduction}
Optical microcavities \cite{Vahala2003} are widely used in optics, for instance, to create good single photon sources \cite{Najer2019, Tomm2021, Vogl2019, Haussler2020}, to image surface layers with very high precision \cite{Mader2015, Benedikter2015, Benedikter2019}, to simulate physical processes like Bose-Einstein condensates \cite{Klaers2010} or non-equilibrium dynamics \cite{Peters2023}, and to perform chemical sensing \cite{Trichet2014, Vallance2016}.
All these examples require many reflections between the mirrors of the cavity, while maintaining strong (transverse) field confinement.
In the past decade, open microcavities have been developed for this purpose \cite{Barbour2011}.
In open microcavities, the two mirrors can be tuned individually to vary the cavity length and lateral mirror displacement.

In order to achieve stronger light-matter interaction, the radius of curvature of the microcavities has been getting smaller \cite{Flatten2018}.
Consequently, the penetration depths can become similar to the cavity length \cite{Babic1992, Koks2021} in such microcavities.
Besides, non-paraxial effects have an increasingly strong influence on the cavity modes \cite{exter2022, koks2022_fine}.
Proper characterization and application of microcavities require precise information about penetration depth and mirror shape.
Planar cavity modes are characterized using emitters in monolithic cavities, where one simultaneously has access to the wavelength spectrum and cavity length-dependent spectrum, where the latter is determined from the angle-resolved emission \cite{Houdre1994, Deng2010}.
Yet, cavity modes in a concave or plano-concave geometry are most commonly characterized at a fixed wavelength \cite{Benedikter2015, Benedikter2019, Uphoff2015, Greuter2014} or a fixed cavity length \cite{Hood2001, weisbuch1992}.
When using emitters in an open microcavity, one also has access to a wavelength and cavity length-dependent emission spectrum for all cavity modes \cite{Riedel2020}.
This has several advantages, as discussed below.

In this paper, we study optical microcavities using broadband, intracavity emitters.
We measure spectra of the fluorescence from these emitters, while slowly tuning the cavity length, resulting in a joined length-wavelength fluorescence spectrum $P(\lambda, L)$.
The key message of this paper is that the analysis of the joined $P(\lambda, L)$ spectrum allows one to determine the penetration and observe mode-mixing in a more controlled way.
First, we distinguish between the different penetration depths \cite{Koks2021} of the cavity mirrors.
Second, we analyze the spectra of the transverse modes and find from the mode spacings that a shape imperfection of the mirror dominates the mode structure.
Last, we zoom in on cavity lengths where the $N=0$ mode is almost frequency degenerate with the $N=6$ mode group. 
We observe an avoided crossing around the wavelength and cavity length where these modes should overlap and quantify it with coupled-mode theory \cite{Paschotta2006, Shemirani2009}.

\section{Setup}
\begin{figure}
    \centering
    \includegraphics[width=\linewidth]{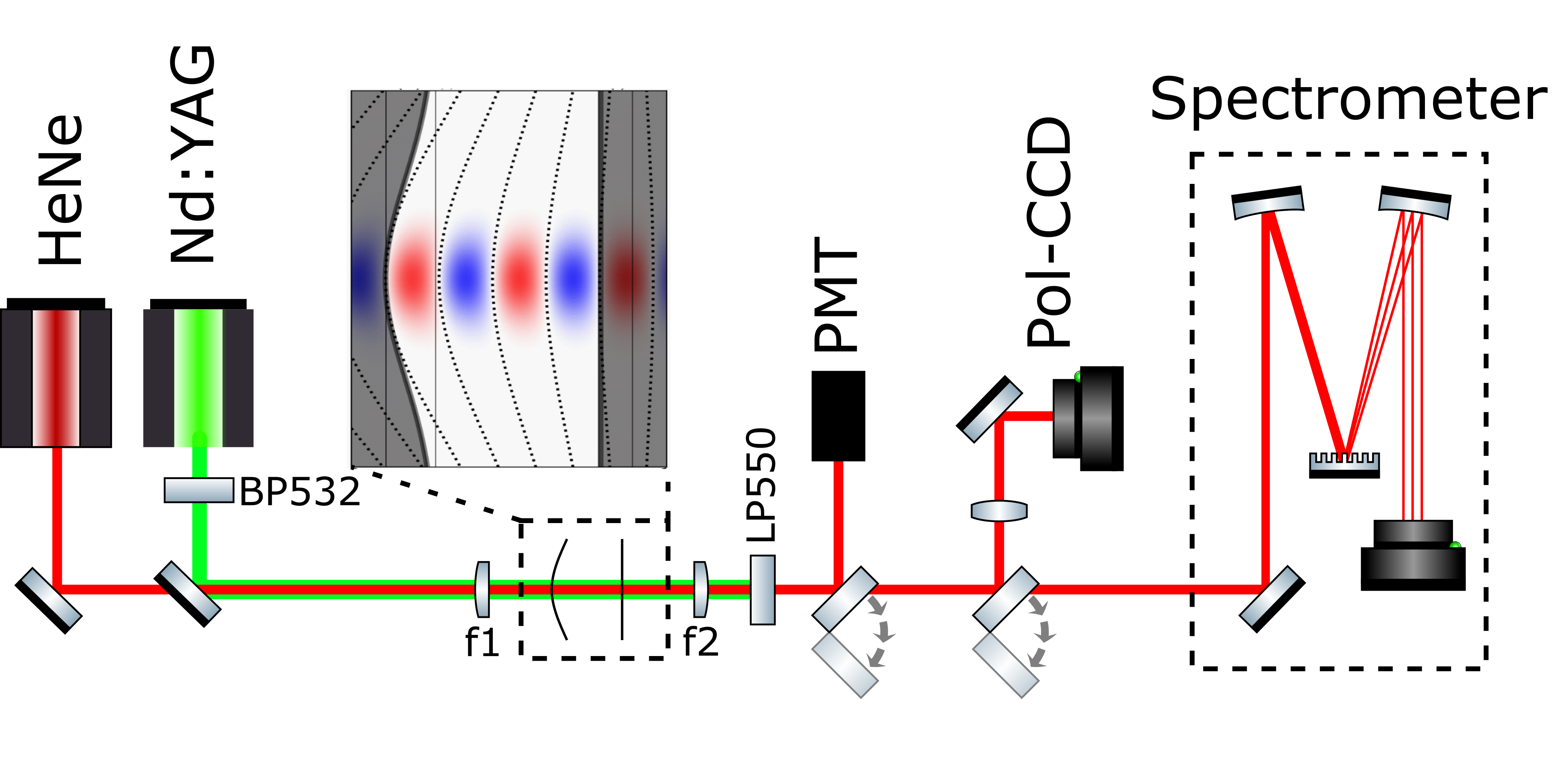}
    \caption{Schematic image of the experimental setup. 
    The microcavity (center + zoom-in) is injected with either a HeNe laser ($\lambda=633$ nm) or a Nd:YAG laser ($\lambda=532$ nm).
    The fluorescence spectra under green off-resonant excitation are observed either with a fiber spectrometer (not shown) or a free-space spectrometer.
    A 10 nm bandpass filter around 532 nm (BP532) is used to filter out fluorescence that may originate from the fiber.
    A 550 nm longpass filter (LP550) is used to block the green laser light transmitted through the cavity.
    }
    \label{fig:experimental setup microcavity nanodiamonds}
\end{figure}

Figure \ref{fig:experimental setup microcavity nanodiamonds} shows a schematic of the experimental setup.
The optical cavity, shown in the center, consists of two highly reflective distributed Bragg reflectors (DBR); one flat and the other curved.
We use an asymmetric set of mirrors, where the flat mirror has a higher transmission ($T=1.8(1) \times 10^{-3}$ at $\lambda=633$ nm, central wavelength $\lambda_c=640$ nm) than the curve mirror ($T=0.3(2) \times 10^{-3}$ at $\lambda=633$ nm, $\lambda_c=610$ nm), such that most of the fluorescent light leaves the microcavity through the flat mirror.
The two mirrors are coated with alternating layers of SiO$_2$ ($n=1.46$) and Ta$_2$O$_5$ ($n=2.09$).
The curved mirror is an H-DBR, meaning that it ends with a high reflective index layer of Ta$_2$O$_5$ to optimize its reflectivity for a given number of layers.
The flat mirror, produced by LaserOptik, is an L-DBR ending with a lower refractive index layer of SiO$_2$, to create a field anti-node close to its surface.
The curved mirror, produced by Oxford HighQ \cite{Trichet2015}, has a small radius of curvature $R=17.7(3)~\mu$m. 
We use a hexapod system to align the mirrors in parallel and tune the cavity lengths $L=$ 3-10 $\mu$m with piezo stacks.

We use nanodiamonds as broadband intracavity emitters.
The nanodiamonds (Adamas Nanotechnologies FND, 40 nm, 1-4 NV$^-$ per nanodiamond) contain NV$^{-}$ and NV$^{0}$ centers whose combined room-temperature emission spectrum (590-700 nm) overlaps with the stopband of our cavity (590-680 nm). 
The nanodiamonds are drop-casted onto the flat mirror. 
Their concentration is not uniform and we scan the flat mirror to find an optimum where the finesse is not strongly degraded and similar to previous works \cite{koks2022_fine}.
More specifically, we study a region where the finesse $F \approx 1000$ instead of the $F \approx 3000$ observed without nanodiamonds. 
By comparing the increased scattering losses of approx. $2 \times 10^{-3}$ with the calculated scattering cross-section 
of the individual nanodiamonds, we estimate to have approximately 1600 nanodiamonds within the mode volume of the microcavity.

In the first experiment, we inject the cavity with a HeNe laser ($\lambda=633$ nm) to probe the finesse and \enquote{length spectrum} of the cavity.
We couple the laser light into the cavity with a $f_1=5$ mm (40X, NA=0.6) lens, and out of the cavity with a $f_2=8$ mm ($\text{NA}=0.5$) aspheric lens.
We rapidly scan the cavity length (scan velocity $\approx 10 ~\mu$m/s) and record the transmission spectrum on a solid-state photomultiplier tube (PMT).
From this scan, we determine a finesse of $F\approx1000$ for the fundamental modes.
Furthermore, we use it to compare the mode structure of a length scan with that of a spectral scan (see below).

In the second set of experiments, which form the bulk of this paper, we measure the fluorescence spectrum from the cavity.
We use a frequency-doubled Nd:YAG-laser ($\lambda=532$ nm) to excite intracavity emitters.
We measure their fluorescence spectrum with two different spectrometers.
The first (fiber-coupled) spectrometer (Ocean Optics QE65000) can image the full fluorescence spectrum with a resolution of $\lambda_{\rm res}=1.2$ nm.
The second (free-space) spectrometer has a smaller spectral range (619-647 nm) but a better spectral resolution of $\lambda_{\rm res}=0.05$ nm, which is comparable to typical linewidths in the cavity fluorescence spectrum.
The photoresponse of this spectrometer is about 1 bitcounts per outcoming cavity photon (the CCD has approx. 8 bitcounts/photon, while the optical transmission is approx. 1/8 from cavity to CCD).
The cavity length is now scanned very slowly (scan velocity $\approx0.3$ nm/s) such that the cavity resonances appear to be quasi-static.
The 1 Hz acquisition rate of the spectrometer is just fast enough to minimize the influence of drift on the measurements.

\section{Joined length-wavelength scan}
\begin{figure*}
    \centering
    \includegraphics[width=\linewidth]{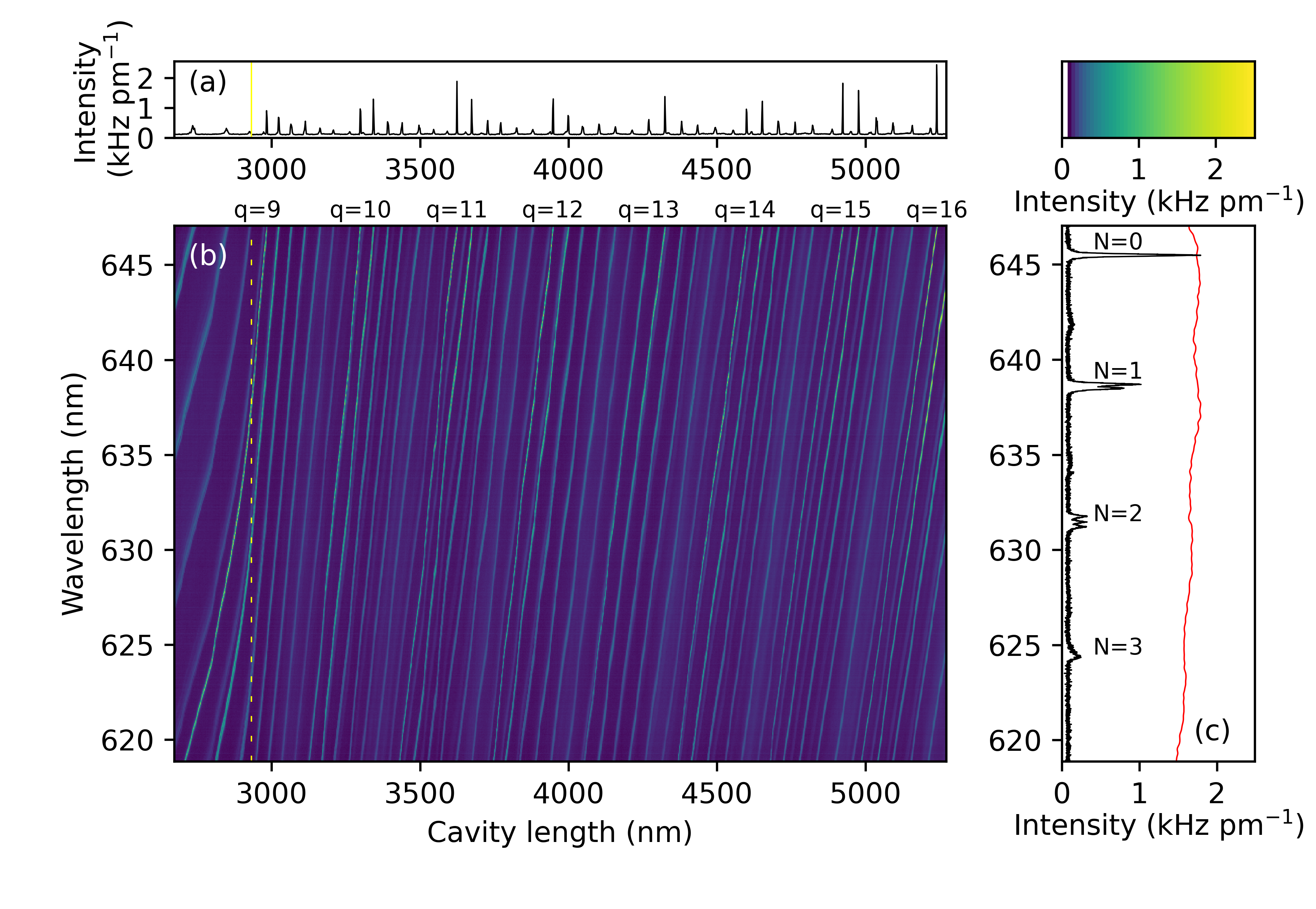}
    \caption{Emission spectrum $P(\lambda, L)$ of NV centers in the optical cavity: (a) horizontal linecut of the $P(\lambda, L)$-map for $\lambda=646$ nm, (b) false-color plot of the fluorescence spectra as a function of the cavity length, (c, black curve) typical spectrum for $q=11$ at $L\approx3.6~\mu$m, (c, red curve) emission spectrum averaged over $L=3.0-5.2~\mu$m.}
    \label{fig:large spectra both cavity length and wavelength}
\end{figure*}

Figure \ref{fig:large spectra both cavity length and wavelength}b shows a typical fluorescence spectrum $P(\lambda, L)$ of the emitters in the cavity. 
This data is measured by slowly increasing the cavity length $L$, while constantly recording the emission spectrum on the free-space spectrometer. 
Each vertical linecut corresponds to a single measurement at fixed $L$.
The linecut for $L\approx3.6~\mu$m is shown as the black curve in Fig. \ref{fig:large spectra both cavity length and wavelength}c
This figure also shows a red curve which is the emission spectrum of the NV centers averaged over a large range of cavity lengths ($L=3.0-5.2~\mu$m).
This shows that the spectrum is relatively uniform over the investigated wavelength range.
The cavity length is estimated by taking a horizontal linecut, plotted in Fig. \ref{fig:large spectra both cavity length and wavelength}a, and fitting the transverse mode splittings to an increasing Gouy phase \cite{Koks2021}. 
The shortest cavity length is $L\approx2.9~\mu$m, indicated by the dashed yellow line, where the substrates of the mirrors almost touch.

The optical modes in Fig. \ref{fig:large spectra both cavity length and wavelength} can be labeled by a longitudinal mode number $q$ and a transverse mode number $N$.
The longitudinal mode number $q$ and cavity length $L$ are estimated based on the mode crossing of the fundamental $N=0$ mode with the $N=6$ mode, where $L+L_{D_1}+L_{D_2}=R\sin(\pi/6)^2=R/4$.
The physical cavity length $L$ is then determined by filling in the values $L_D$ and $R$ from the fits in Fig. \ref{fig:penetration depths plot} below.
The number of half-wavelengths that fit in this cavity length results in a longitudinal mode number $q=2L/\lambda=13.1(3)$ for the fundamental mode at this crossing at $\lambda=637$ nm.
The large uncertainty in $q$ is due to the difficulty to determine the radius of curvature and modal penetration depth for all longitudinal mode numbers simultaneously.
For a combination of a H-DBR and L-DBR, we expect $q$ to be a half-integer \cite{Koks2021}. 
However, due to the asymmetry of the mirror stopbands and slightly thinner final layer of the L-DBR, the reflection phase of the L-DBR is non-zero, which results in an unconventional value of $q$.
For labeling purposes, we round $q$ to integer values. 
The exact value of $q$ has no influence on the measurements shown below, but will be discussed in more detail in the Appendix.


\section{Penetration depths}
\begin{figure*} 
    \includegraphics[width=\linewidth]{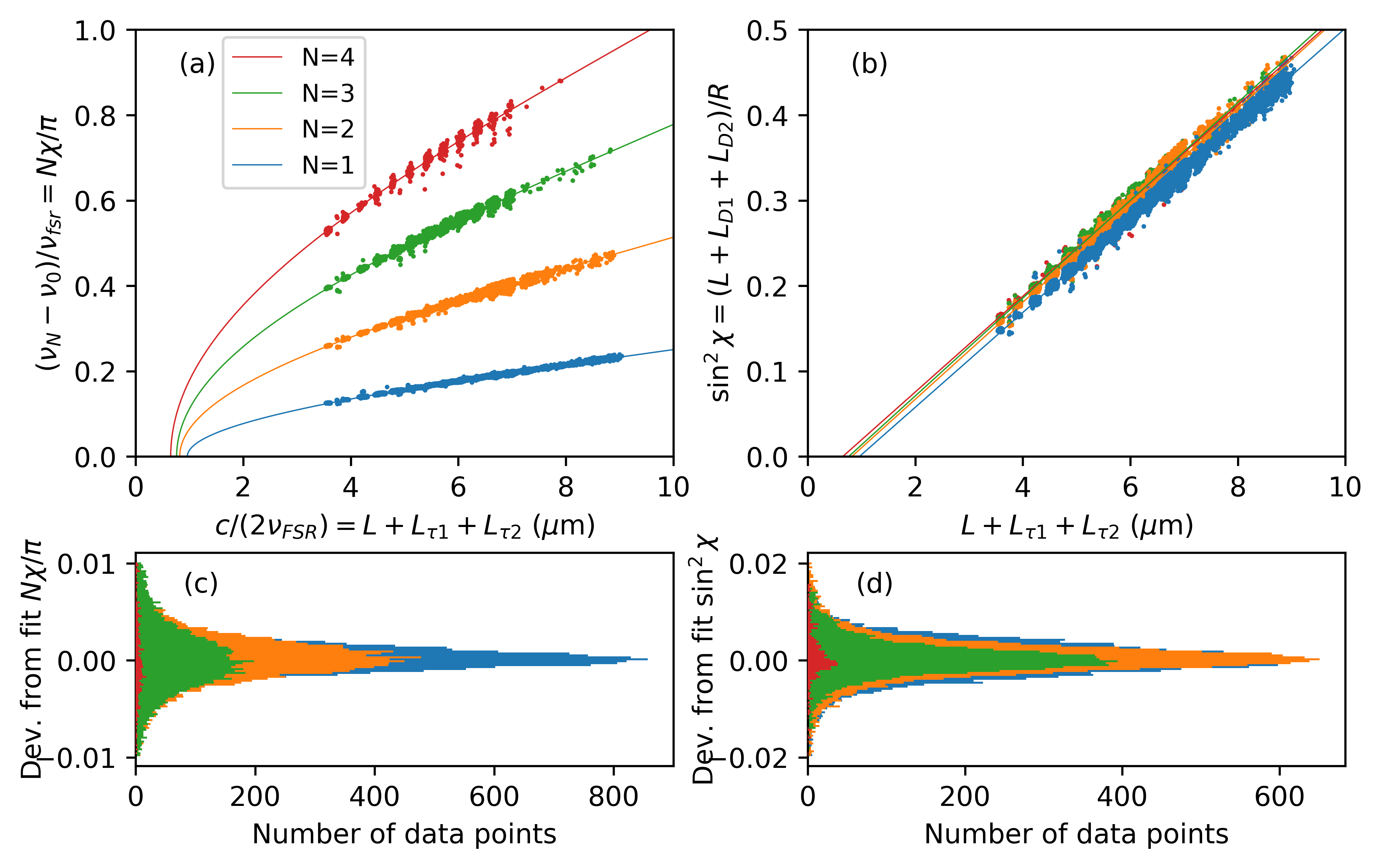}
    \caption{
    (a) The transverse mode splitting for the first 4 transverse mode groups as a function of the inverse of the free spectral range. (b) The Gouy phase converted to a \enquote{modal cavity length} such that both axes depend linearly on cavity length. The crossing with the horizontal axis yields the difference between the penetration depths. (c-d) Histograms of the deviation from the fits for the vertical axis. The horizontal axis shows the number of data points in a bin of 0.0002(0.0004) for c(d).
    }
    \label{fig:penetration depths plot}
\end{figure*}

The frequency- and angle-dependent reflection of any DBR can be described by three penetration depths: $L_\varphi$, $L_\tau$ and $L_D$ \cite{Babic1992, Babic1993, Koks2021}.
The phase penetration depth $L_\varphi$ is the shift of the (anti-)node when the frequency $\nu$ is detuned from the DBR's central frequency $\nu_c$.
The frequency penetration depth $L_\tau$ is the extra length required to correct for the time delay of a reflected laser pulse.
The modal penetration depth $L_D$ describes the shift of the focus of a converging beam inside the mirror.
The penetration depths are different for H-DBRs and L-DBRs \cite{Koks2021, Babic1992}:
\begin{equation}
    L_\tau=\frac{1}{n_H-n_L} \frac{\lambda_c}{4} ~\text{(H-DBR)}
\end{equation}
and
\begin{equation}
    L_\tau=\frac{n_H n_L}{n_H-n_L} \frac{\lambda_c}{4}~ \text{(L-DBR)}
\end{equation}
where $n_H$ and $n_L$ are the high and low refractive indices of the DBR pairs.
The modal penetration is related to the frequency penetration depth \cite{Koks2021, Babic1993}, as
\begin{equation}
    L_D=\frac{L_\tau}{2} \left(\frac{1}{n_H^2}+\frac{1}{n_L^2}\right).
\end{equation}
Both equations assume that the DBR is illuminated from air ($n=1$).
The phase penetration depths is frequency dependent, $L_\varphi=\frac{\nu-\nu_c}{\nu}L_{\tau}$, where $\nu_c$ is the central frequency of the stopband.
The penetration depths of the two different mirrors used here are referred to as $L_{\tau,1},L_{\tau,2}, L_{D,1}, L_{D,2}$.

The frequency spacing between the cavity modes depends on different penetration depths.
This can be seen from the cavity resonance condition \cite{Koks2021},
\begin{equation}\label{eq:resonance condiction in main text}
    \frac{2L}{c} \nu + \frac{2(L_{\tau,1}+L_{\tau,2})}{c} (\nu-\nu_c)= q + (N+1)\frac{\chi}{\pi},
\end{equation}
where $L$ is the physical (on-axis) distance between the two mirrors, $c$ is the speed of light, and $\chi$ is the Gouy phase (see below).
The frequency difference between two consecutive fundamental modes at fixed $L$, the so-called free spectral range, depends only on the frequency penetration depth,
\begin{equation}
    \nu_{q+1,0}-\nu_{q,0}=\nu_{\rm fsr}=\frac{c}{2(L+L_{\tau,1}+L_{\tau,2})}
\end{equation}
and therefore,
\begin{equation}
    L+L_{\tau,1}+L_{\tau,2} = \frac{c}{2 (\nu_{q+1,0}-\nu_{q,0})}. 
    \label{eq:resonance only frequency penetration depth}
\end{equation}
The transverse mode spacings and the associated Gouy phase $\chi$, on the other hand, depend only on the modal penetration depth \cite{Koks2021} via 
\begin{equation}
    \pi \frac{\nu_{q,N}-\nu_{q,0}}{N \nu_{\rm fsr}}=\chi=\arcsin\sqrt{\frac{L+L_{D,1}+L_{D,2}}{R}}
\end{equation}
and therefore, 
\begin{equation}
    \frac{L+L_{D,1}+L_{D,2}}{R}= \sin^2(\frac{\pi}{N} \frac{\nu_{q,N}-\nu_{q,0}}{\nu_{q+1,0}-\nu_{q,0}}).
    \label{eq:resonance only modal length}
\end{equation}
From the difference between these quantities, we find the difference between the penetration depths of the mirror pair $L_{\tau,1}+L_{\tau,2}- L_{D,1}- L_{D,2}$, without having to rely on an absolute longitudinal mode number $q$.

Figure \ref{fig:penetration depths plot} shows our procedure to determine the two different penetration depths.
These were measured using the fiber spectrometer with the larger spectral range.
We only use cavity modes between $\lambda=610-670$ nm such that a non-linearity in the reflection phase of the DBR is relatively small \cite{Hood2001}. 
The data in the figure are corrected for this 3$^{\rm rd}$-order non-linearity (corrections are $\leq$20\%, see Appendix \ref{sec:appendix penetration depth}).
The horizontal axis indicates the cavity length, as determined by the inverse of the free spectral range. 
The vertical axis shows the transverse mode splittings, normalized to the free spectral range.
Figures \ref{fig:penetration depths plot}c and \ref{fig:penetration depths plot}d show the spread in the data points.
This is mainly due to the modest resolution of the fiber-spectrometer of 1.2 nm.
The relative resolution is worse at larger cavity lengths because the free spectral range by which the data is normalized becomes smaller.
Furthermore, the higher N groups show an increased spread due to the spectral splittings, also shown in Fig. \ref{fig:fine structure}. 
Note that the periodic structure is an artifact of the shift of the free spectral range out of the cavity mirrors' stopband.


The difference in penetration depths $L_\tau-L_D$ is visible as the crossing with the x-axis in Fig. \ref{fig:penetration depths plot}.
This can be seen best in Fig. \ref{fig:penetration depths plot}b.
A linear fit of Fig. \ref{fig:penetration depths plot}b gives penetration depths for transverse modes $N=1-4$ of $(L_{\tau,1}+L_{\tau,2}- L_{D,1}- L_{D,2})/2= 0.481(2), 0.410(2), 0.381(3), 0.33(1)~\mu$m.
The radii of curvature are $R=18.02(1), 17.60(1), 17.45(2), 17.84(9)~\mu$m.
From theory \cite{Koks2021} we expect that the H-DBR (curved mirror) has a penetration depth $L_\tau=0.25~\mu$m, and the L-DBR has a penetration depth $L_\tau=0.77~\mu$m.
This means that the theoretical difference in penetration depths $(L_{\tau,1}+L_{\tau,2}- L_{D,1}- L_{D,2})/2=0.33$ $\mu$m.
The frequency penetration depth for the modified (top layer $0.8\times \lambda_c/(4 n_L)$) L-DBR will be smaller than for a standard L-DBR.
From the slope of the reflection phase versus wavelength data provided by the manufacturer, we find a theoretical value for the L-DBR $L_\tau=0.69(2)~\mu$m, which results in a slightly smaller theoretical prediction $(L_{\tau,1}+L_{\tau,2}- L_{D,1}- L_{D,2})/2=0.30(1)~\mu$m.
The measured values of $L_\tau-L_D$ agree reasonably well with theoretical predictions.
Still, the measured values are somewhat larger than predicted.
This might be due to the asymmetric mirror set with different central wavelengths. 
The data also suggests $N$-dependent variations in curvature and penetration depths, possibly due to mirror shape effects \cite{Koks2022}, where each transverse mode scans a different part of the curved mirror.
The error bars of the estimated (differences in) penetration depths underestimate the influence of noise, as they are computer-generated and hence based on intrinsic estimates of the noise. 
Still, the trend of these estimates with transverse order $N$ is undeniable.

\section{Transverse mode group structure } 

\begin{figure}
    \centering
    \includegraphics[width=\linewidth]{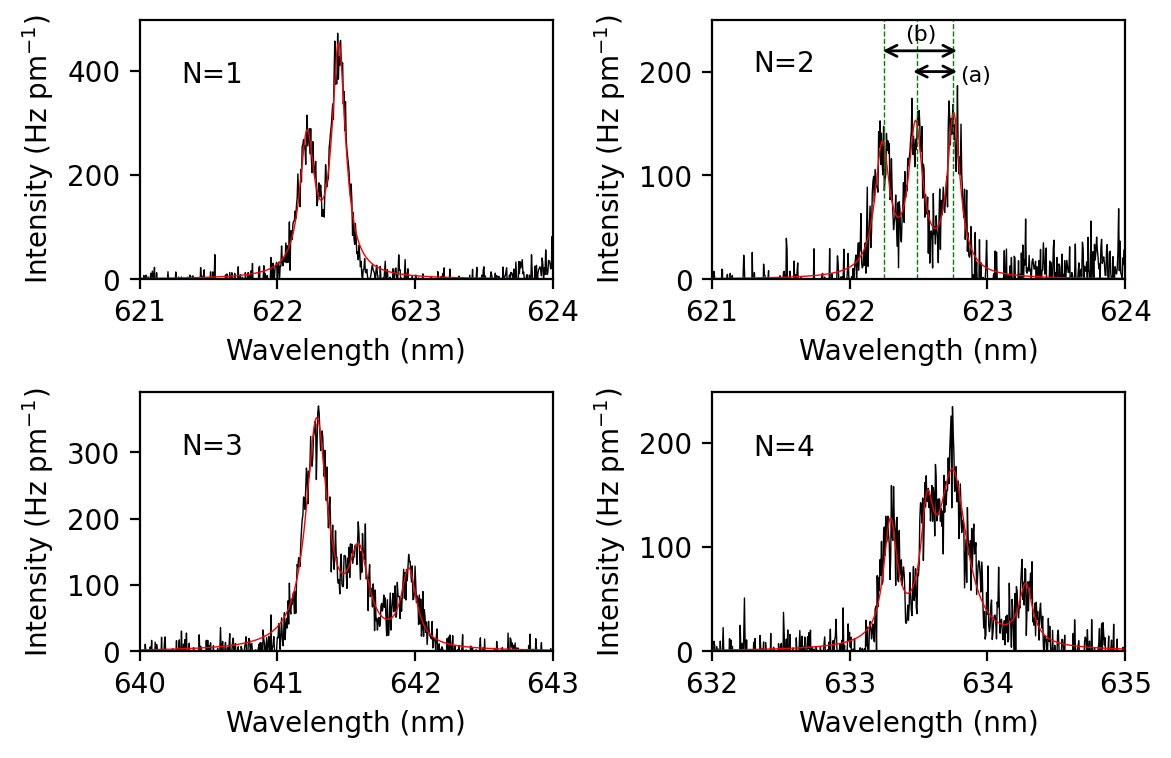}
    \caption{Mode structure for transverse mode groups $N=1-4$. 
    Each transverse mode group is fitted with 2, 3, 3 and 4 Lorentzians for the $N=1, 2, 3$ and $4$, respectively. 
    The $N=2$ plot show peak distances that correspond to the (a) and (b) labels in Table \ref{tab: fine structure values}.
    }
    \label{fig:fine structure}
\end{figure}

Figure \ref{fig:fine structure} shows zoomed-in spectra of the transverse mode groups $N=1$ to $N=4$. 
The $N=1$ and $N=2$ groups were measured at $q=11$, where the modes are least coupled to any other transverse mode groups.
The $N=3$ and $N=4$ transverse groups were measured at $q=8$ where the two mirrors almost touch, such that vibrations are smaller and the peaks become better resolvable.
Each transverse mode group $N$ is predicted to consist of $N+1$ separate peaks. 
The observed number of peaks in the $N=3$ and $N=4$ spectra is $N$ instead of the expected $N+1$, presumably because two peaks overlap.

\begin{table}[H]

\caption{Mode structure, relative to the free spectral range, determined from (i) emission spectra at fixed cavity length, (ii) length scans at fixed wavelength ($\lambda=633$ nm), and (iii) theoretical values for a perfectly spherical mirror.
The $N=1$ structure contains two peaks, while the $N=2$ structure contains three peaks.
The two relative splittings for $N=2$ are (a) and (b) as indicated in Fig. \ref{fig:fine structure}.
}
\begin{center}\label{tab: fine structure values}
\begin{tabular}{ |c|c|c|c|c| } 
\hline
\quad & $\Delta \lambda/\lambda_{\rm fsr}$ & $\Delta L/(\lambda/2)$ & $\Delta\varphi_{th}/(2\pi)$ \\
\hline 
N=1 & $5.35(2) \times10^{-3}$ & $5.1(6)\times 10^{-3}$ & $1.78\times 10^{-3}$ \\ 
N=2 (a)& $5.63(5)\times 10^{-3}$ & $6.6(6)\times 10^{-3}$ & $2.78\times 10^{-3}$\\ 
N=2 (b)& $12.05(5)\times 10^{-3}$ & $13(1)\times 10^{-3}$ & $3.56\times 10^{-3}$\\ 
\hline
\end{tabular}
\end{center}

\end{table}

The measured transverse group structure is quantified by determining the distance between the peaks.
In the case of a perfectly spherical mirror, the mode structure is determined by a fine structure, $\Delta \lambda/\lambda_{\rm fsr}=(\ell \cdot s + 3 \ell^2/8)/(2\pi k R)$, where $\ell$ and $s$ are the angular momentum and spin polarization of the mode \cite{koks2022_fine, exter2022}.
For the $N=1$ and $N=2$ mode groups, we observe the expected number of peaks, so we can make a comparison between theory and measurement.
Table \ref{tab: fine structure values} shows this comparison, where we also added the splitting from the mode structure measured with a length scan using a HeNe laser.

The theoretically predicted values for the non-paraxial fine structure are a factor 2-3 smaller than the measured values.
We thus conclude that technical effects dominate over the intrinsic non-paraxial effects. 
This is most likely due to astigmatism in the curved mirror, which shifts the fundamental fine structure towards more equidistant peaks, as observed for $N=2$ \cite{koks2022_fine}.
Another explanation of the astigmatism could be a symmetry breaking due to a gradient concentration of nanodiamonds.
We also observe that the peaks in Fig. \ref{fig:fine structure} have different integrated powers. 
However, there is no clear trend visible where, for example, a central peak with a more centered mode is stronger than side peaks with a more spread-out mode.
Hence, we can not conclude on a clear difference in the concentration of the nanodiamonds. 

The advantage of a spectral measurement over a length scan is that it hardly depends on the incoupling of light \cite{koks2022_fine}.
Furthermore, it is not influenced by fluctuations in the piezo-velocity, such that the uncertainties are smaller.
The mode structure is similar to the spectral and cavity length scan.

The disadvantage of the described $P(\lambda,L)$ method, however, is the limited operation of the finesse. 
In particular, cavities with very high finesse \cite{Jin2022, Wachter2019} will suffer significantly from the additional scattering losses. 
Although we try to stay in a regime where the finesse hardly suffers, we do observe additional scattering losses of the order $2 \times 10^{-3}$, corresponding to approximately 1600 nanodiamonds.


\section{Mode coupling}

\begin{figure}[H] 
    \centering
    \includegraphics[width=\linewidth]{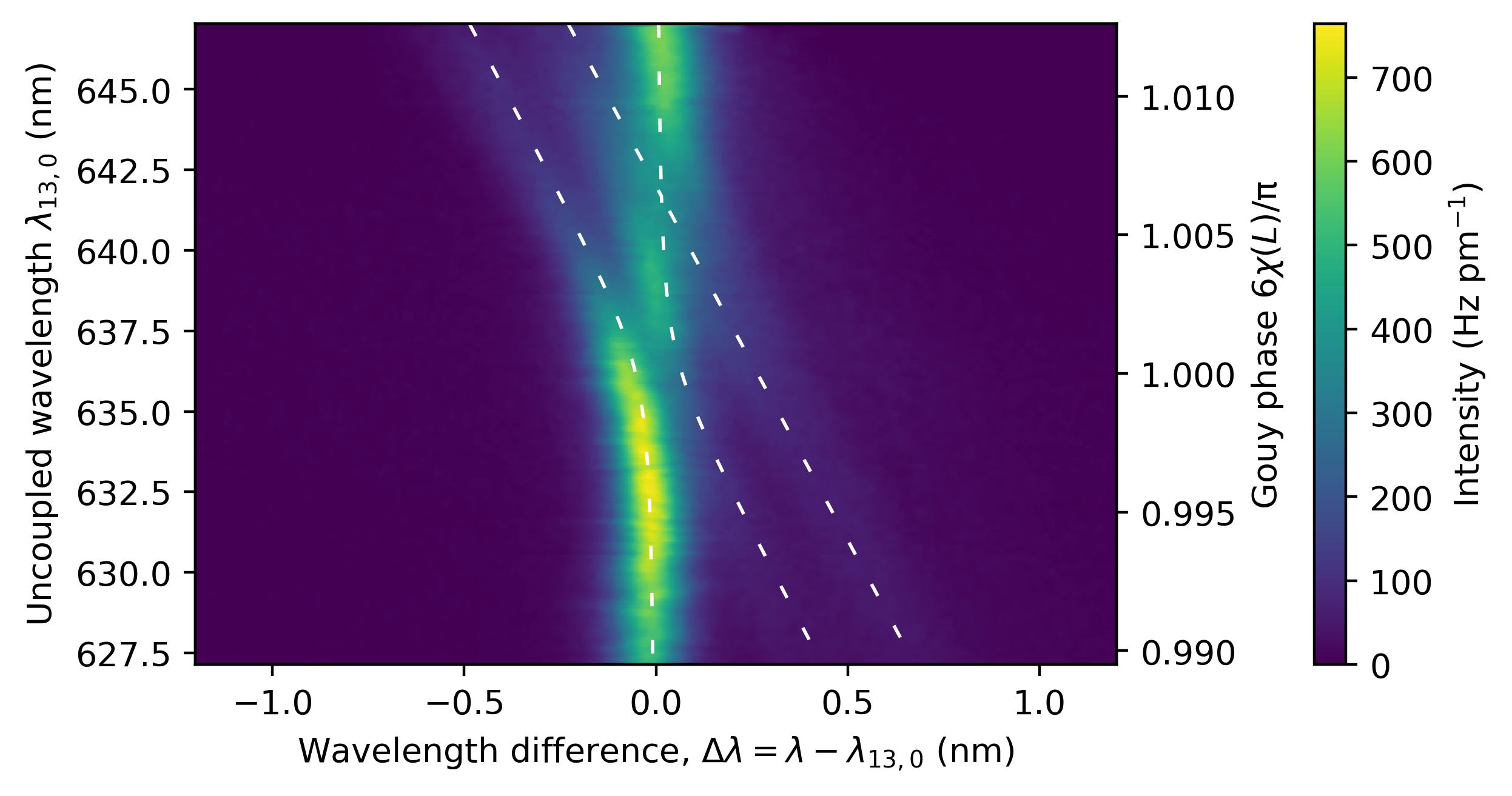}
    \caption{False color plot of the fluorescence in the $N=0$ and $N=6$ modes around their frequency degeneracy. 
    The vertical axis indicates the cavity-length-dependent wavelength $\lambda_{13,0}$ at which the uncoupled $N=0$ ($q=13$) mode is resonant. 
    The horizontal axis indicates the spectral distance $\lambda-\lambda_{13,0}$ from this uncoupled $N=0$ mode. 
    The white curves follow a theoretical model of coupled modes (see text for details). 
    }
    \label{fig:mode coupling falsecolorplot}
\end{figure}

Figure \ref{fig:mode coupling falsecolorplot} shows a false color plot of $P(\lambda, L)$ in the region where the $(q=13, N=0)$ and $(q=12, N=6)$ modes cross.
This is a zoom-in from figure \ref{fig:large spectra both cavity length and wavelength}b around the $q=13$ longitudinal mode, but the horizontal and vertical axes are interchanged and rescaled.
The vertical axis now shows the wavelength $\lambda_{13, 0}$ at which an uncoupled $(q=13, N=0)$ mode would be resonant.
The increasing value of $\lambda_{13, 0}$ corresponds to an increasing cavity length in the measurements, which can also be described by an increasing Gouy phase, as shown on the (second) y-axis on the right.
The horizontal axis shows the spectrum in terms of wavelength detuning $\lambda-\lambda_{13, 0}$.
Comparable figures in refs. \cite{Benedikter2015, Trichet2018} present data as a function of longitudinal mode number $q$, i.e. a $P(q, L)$ map is shown instead of a $P(\lambda, L)$ map.
The advantage of a $P(\lambda, L)$ map is that $\lambda$ is a continuous variable and not an integer like $q$.
This is especially important for small cavities, where the Gouy phase $\chi$ changes rapidly with cavity length and the steps due to discrete changes in $q$ are too large to accurately measure coupling in the length spectra \cite{Koks2022}.

The accurate determination of $\lambda_{13,0}$ required some refinements of the data.
First, we shifted all measured spectra such that the intensity-weighted wavelength around the $N=0$ and $N=6$ modes was set to zero. 
From the resulting image, a straight line could be drawn for the $N=0$ mode between the highest and lowest wavelengths, where the $N=0$ and $N=6$ modes are not coupled anymore (see top and bottom of Fig. \ref{fig:mode coupling falsecolorplot}).
This straight line yields the values $\lambda_{13,0}$ in Fig. \ref{fig:mode coupling falsecolorplot}.
A Gaussian filter was used to make the $N=6$ modes (visible as slanted lines) better visible.
Without these corrections, the coupling is still visible but less clean. 

We can quantify this avoided crossing using a coupled-mode model with a coupling matrix \cite{Koks2022}
\begin{equation}
\begin{pmatrix}
0 & -M \\ 
-M& \Delta\varphi
\end{pmatrix},    
\end{equation}
where $\Delta\varphi=6\chi-\pi$ is the one-way phase difference between the $N=6$ and $N=0$ modes, and $M$ is the coupling parameter.
Using a \enquote{fit-by-eye} we find $M=0.014(3)$.
This value is comparable to the coupling parameter $M=0.016(2)$ found in a similar cavity \cite{Koks2022}.
This shows the equivalence between measurements in the frequency-domain and the far-field.
Previous measurements on a similar cavity \cite{Koks2022} showed that the origin of mode-mixing was a slight deviation of the curved (H-DBR) mirror shape. 
An AFM topography on that mirror showed a defect at the mirror center which gave the cavity a slight bathtub shape. 
The curved mirror that is used in the current experiments is produced in the same way \cite{Trichet2015} and, although we have not performed AFM measurements on this mirror, we believe the origin to be the same. 


\section{Conclusions}

We have demonstrated the use of broadband intracavity emitters for the characterization of microcavities.
The obtained broadband spectra $P(\lambda, L)$ enable accurate determination of the difference between two DBR penetration depths $L_\tau-L_D$, in reasonable agreement with theoretical expectations.
Furthermore, mode-coupling can be measured without having to switch between discrete values of $q$.
This is crucial in very small cavities, where the Gouy phase changes rapidly with cavity length. 
Although the measurements could also be done using a tunable laser or white light source, the advantages of using emitters in a cavity are that they are typically already in place inside the cavity \cite{Riedel2020} and that the light source does not depend on incoupling. In many experiments, the experiments shown above can thus be done without much additional effort.

The description of mode formation in a microcavity becomes intriguing when microcavities become smaller.
For small radii of curvature, the penetration depth can become significant, especially when L-DBRs are used.
Indeed, the measured difference between two penetration depths $L_\tau-L_D$ is approximately half a wavelength, so an erroneous determination of the penetration depth can easily result in a wrong estimation of the longitudinal mode number $q$.
The method that is described here is better than a previously presented method \cite{Koks2021} because it does not rely on an exact cavity length $L$ or longitudinal mode number $q$. 
Furthermore, the penetration depth can be determined for each individual transverse mode group.
Our measurements show a difference in penetration depth for the different transverse mode groups, possibly due to the mirror shape that is probed differently for each transverse mode.
This mirror shape also causes mode-mixing, which can be observed as an avoided crossing of two modes.
The full $P(\lambda, L)$ map of the microcavity reveals all these aspects in a single picture.

\section{Acknowledgments}
We thank A. Trichet from Oxford HighQ for providing the curved mirrors.

\bibliographystyle{apsrev4-1}
\bibliography{bib}


\section{Third-order frequency dispersion of DBRs}
\label{sec:appendix penetration depth}
This Appendix describes the influence of the third-order frequency dispersion of DBRs on the measurements.
For most cases, the reflection phase of the DBR is linearly approximated, $\varphi=\frac{2 L_\tau}{c} (\nu-\nu_c)$, with the central frequency $\nu_c$ and frequency penetration depth $L_\tau$.
If the frequency is near the edge of the DBR stopband, this relation becomes non-linear \cite{Hood2001, Thorpe2005}.
This has consequences for the accurate determination of the penetration depths.

The complex reflection amplitude of a DBR can be derived from coupled mode theory \cite{Haus, Yeh}.
Its frequency dependence can be derived from an expansion of the phase in the reflection amplitude $\Gamma_0=|\Gamma_0| e^{i\varphi}$ \cite{Koks2021}, where $|\Gamma_0|=1$. The result
\begin{equation}
    \varphi=\arcsin[\tau_0 (\omega-\omega_c)]\approx \tau_0(\omega-\omega_c)+\frac{\tau_0^3 (\omega-\omega_c)^3}{6} + ... ~,
\end{equation}
with $\tau_0 \omega_c=\pi n_H/(n_H-n_L)$, is only valid when the material outside of the DBR has refractive index $n_{in}=n_H$ or $n_{in}=n_L$.
For the general case, we need to consider one more reflection from its first layer, such that the new reflection amplitude is
\begin{equation}\label{eq:DBR:withouttoplayer}
    \Gamma=-\frac{r\mp\Gamma_0 e^{i2(\omega-\omega_c)\Delta\tau}}{1\pm r \Gamma_0 e^{i2(\omega-\omega_c)\Delta\tau}}.
\end{equation}
where the $+(-)$-sign is used for an initial layer with low(high) refractive index and $\Delta\tau=\pi/(2\omega_c)$ is the transit time through the first layer.
The reflection from the first interface is $r=\frac{n_{in}-n_{L/H}}{n_{in}+n_{L/H}}$, where the environment has refractive index refractive index $n_{in}$. Our experiments are conducted in air, so we can set $n_{in}=1$. 
The approximate reflection phase of the mirror including this top layer is
\begin{equation}\label{eq:DBR:withtoplayer}
    \Gamma=|\Gamma|e^{i(\tau_{L/H} (\omega-\omega_c)+\mu_{L/H}(\omega-\omega_c)^3/\omega_c^3)},
\end{equation}
where $\tau_{L/H}$ and $\mu_{L/H}$ are constants which are determined from the refractive indices of the DBR. 
The frequency penetration depth is related through $L_{L/H}=\frac{ \tau_{L/H}}{2 n_{in}}c$.
If we assume that $|\Gamma_0|=|\Gamma|=1$ and compare the first and third derivatives of equations (\ref{eq:DBR:withouttoplayer}) and (\ref{eq:DBR:withtoplayer}), as was done in \cite{Babic1992}, we find for the H-DBR,
\begin{equation}
    \omega_c\tau_H=\pi \frac{n_{in}}{n_H}\frac{n_H}{n_H-n_L} \quad ~~~~ \mu_H = \pi  \frac{n_{in}}{6 n_H}\left(\frac{n_H}{n_H-n_L}\right)^3
\end{equation}
and for the L-DBR
\begin{equation}\label{eq:DBR nonlinearity}
    \omega_c\tau_L=\pi \frac{n_{L}}{n_{in}}\frac{n_H}{n_H-n_L} \quad ~~~~ \mu_L  = \pi \frac{n_{L}}{6 n_{in}}\left(\frac{n_H}{n_H-n_L}\right)^3.
\end{equation}
For our mirrors, with $n_L=1.46$, $n_H=2.09$ and the refractive index of our enviroment $n_{in}=1$, we expect $\mu_H=9.1$ and $\mu_L=27.9$.

The resonance condition of the cavity, including the third-order non-linearity of the DBR is 
\begin{multline}
    \label{eq:appendix resonance condition}
    \nu\frac{2L}{c}
    =q+(N+1)\chi/\pi\\-\frac{2(L_{\tau,1}+L_{\tau,2})}{c}(\nu-\nu_c)-\mu\frac{(\nu-\nu_c)^3}{\nu_c^3},
\end{multline}
which is an extended version of Eq. (\ref{eq:resonance condiction in main text}) in the main text.
$\mu=\mu_1+\mu_2$ is the combined non-linearity constant of both mirrors.
The free spectral range $\nu_{\rm fsr}=\nu_{q+1,0}-\nu_{q,0}$ in this case is given by;
\begin{widetext}
\begin{equation}
            \nu_{\rm fsr}\frac{2(L+L_{\tau,1}+L_{\tau,2})}{c}=1- \mu\frac{(\nu_{q+1,0}-\nu_c)^3-(\nu_{q,0}-\nu_c)^3}{\nu_c^3}\approx 1-3\mu \frac{\nu_{\rm fsr}}{\nu_c}  \left( \frac{\nu-\nu_c}{\nu_c}\right)^2,
\end{equation}
\end{widetext}
which is an extended version of Eq. (\ref{eq:resonance only frequency penetration depth}) in the main text.
We can express this in terms of an effective cavity length $L_{\rm eff}$ normalized by the averaged frequency $\overline{\nu}=(\nu_{q+1,0}+\nu_{q,0})/2$ 
\begin{widetext}
\begin{equation}\label{eq:dbr final equation}
    \frac{\overline{\nu}}{\nu_{\rm fsr}}=\frac{L_{\rm eff}}{\lambda/2}\approx
    q+\frac{1}{2}+\frac{2(L_{\tau,1}+L_{\tau,2})}{c}\nu_c+3\mu \left( \frac{\overline{\nu}-\nu_c}{\nu_c}\right)^2 \frac{\overline{\nu}}{\nu_c},
    \end{equation}
\end{widetext}
where $q+1/2$ is the average $q$ of two subsequent fundamental modes, $\frac{2 (L_{\tau,1}+L_{\tau,2})}{c}\nu_c$ originates from the linear frequency dispersion, and the final term is a third-order correction to the frequency penetration depth.

\begin{figure}[H]
    \centering
    \includegraphics[width=\linewidth]{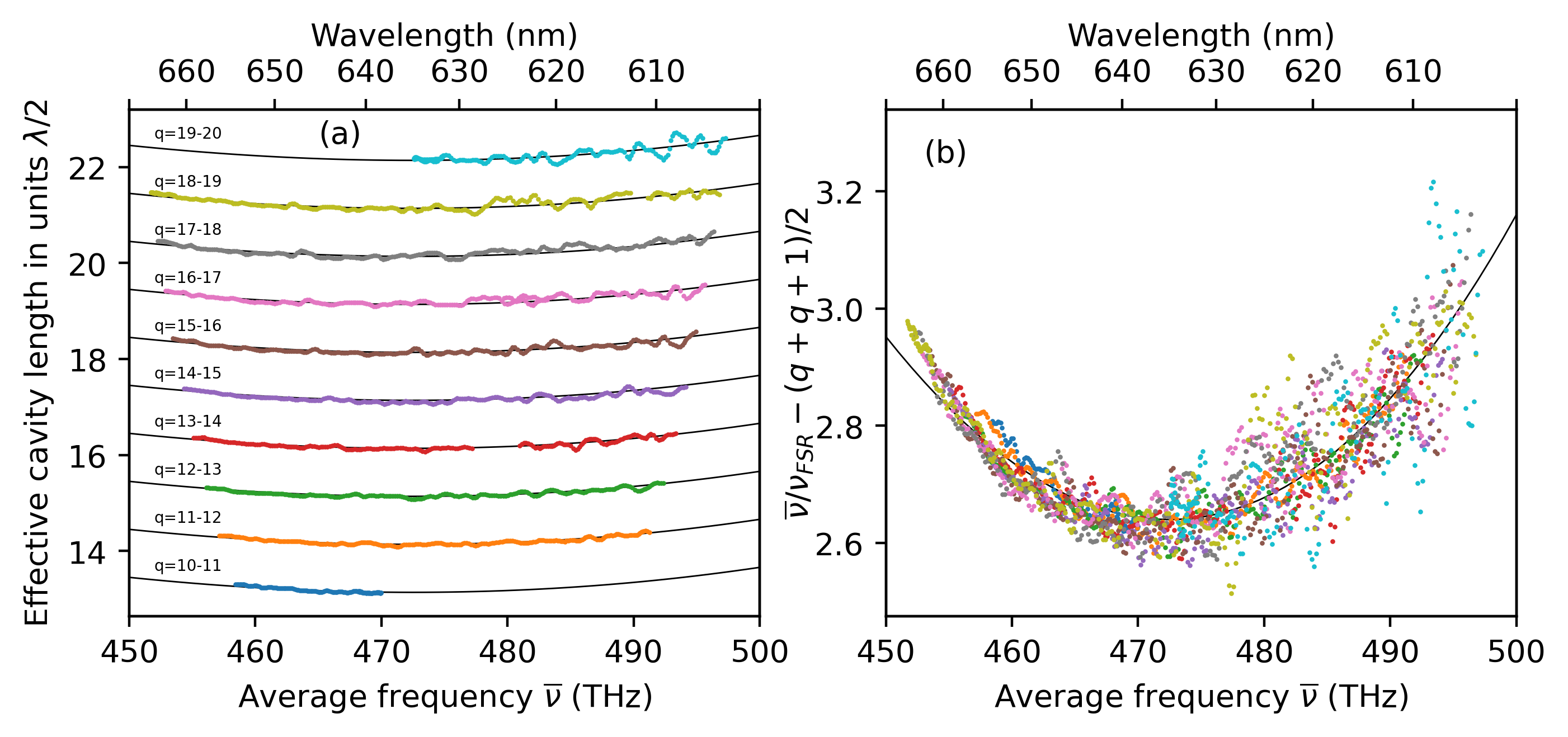}
    \caption{Effective cavity length, $L_{\rm eff}$, calculated from the free spectral range between the $N=0$ modes with $q$ and $q+1$, where 
    $q=10-19$ are indicated with different colors. 
    The points at a higher average frequency are more noisy because the intracavity NV centers are less bright. (a) Original data, (b) all data shifted by $q+1/2$, where $q$ is the lowest number in the labels $q=10-11$, $q=11-12$ etc. in (a).
    }
    \label{fig:penetration depth distance fit}
\end{figure}

Figure \ref{fig:penetration depth distance fit}a shows the observed effective cavity lengths, expressed in units $\lambda/2$.
Each curve is one longitudinal mode number higher, as expected.
Furthermore, the parabolic trend from equation (\ref{eq:dbr final equation}) is visible.
This parabola is better visible in Fig. \ref{fig:penetration depth distance fit}b, where all curves are shifted by $q+1/2$ and now lie on top of each other.
A simultaneous fit of all data with Eq. (\ref{eq:dbr final equation}), yields a non-linearity factor $\mu=48.3(6)$ and a central frequency $\lambda_c=635(1)$ nm ($\nu_c=472.5(8)$ THz).
The measured value for $\mu$ agrees reasonably well with the theoretically predicted value of $\mu=\mu_H+\mu_L=37.0$. 
But the central wavelength is surprisingly close to the central wavelength of the H-DBR, $\lambda_c=640$ nm ($\nu_c=469$ THz).
From theory, one would rather expect the central wavelength to be closer to the L-DBR, $\lambda_c=610$ nm ($\nu_c=492$ THz), since its value for $\mu_L$ is larger than $\mu_H$.
This discrepancy from theory might be due to the large spread in data points near 610 nm, where the NV centers are less bright.

The offset of the parabola in Fig. \ref{fig:penetration depth distance fit}b contains information about $L_\tau$.
The figure shows the effective cavity length $\overline{\nu}/\nu_{\rm fsr}$, divided by $\lambda/2$ and shifted by $q+1/2$.
In contrast to the main text, the exact value of $q$ is now important.
The $q$ that is subtracted is the same integer value that is used for labeling in the main text and in Fig. \ref{fig:penetration depth distance fit}, but this is not entirely correct.
For the combination of an ideal L-DBR and H-DBR, we expect $q$ to be half-integer \cite{Koks2021}, but for our L-DBR, with a top layer of $0.8\times\lambda_c/(4 n_L)$ instead of $\lambda_c/(4 n_L)$, we expect a slightly different $q$.
The best experimental estimate of the longitudinal mode number is $q=13.1(3)$ given in the main text, which is thus 0.1(3) more than the integer labels used to convert Fig. \ref{fig:penetration depth distance fit}a to Fig. \ref{fig:penetration depth distance fit}b.
Therefore, if we want to interpret the vertical axes of Fig. \ref{fig:penetration depth distance fit}b correctly, we should subtract an additional 0.1(3), which we then use to roughly estimate the penetration depth from the offset value in Fig. \ref{fig:penetration depth distance fit}b to be  $\frac{2 (L_{\tau,1}+L_{\tau,2})}{c}\nu_c\approx2.5(3)$ or $L_{\tau,1}+L_{\tau,2}\approx0.8(1)~\mu$m. This is somewhat smaller than the theoretical prediction $\frac{2(L_{\tau,1,th}+L_{\tau,2,th})}{c}\nu_c=3.20$ or $L_{\tau,1,th}+L_{\tau,2,th}=1.02~\mu$m ($L_{\tau,1,th}=0.77$ and $L_{\tau,2,th}=0.25$).
The measurements in the main text do not rely on $q$ and are thus a more reliable measurement of the penetration depth.

\end{document}